# Counterfactual Entanglement and Nonlocal Correlations in Separable States

Oliver Cohen

Theoretical Physics Research Unit, Birkbeck College, University of London,

Malet Street, London WC1E 7HX, UK

and Department of Physics, Carnegie Mellon University, Pittsburgh, PA 15213, USA

e-mail: ocohen@andrew.cmu.edu

**ABSTRACT**

It is shown that the outcomes of measurements on systems in separable mixed states can be partitioned, via subsequent measurements on a disentangled extraneous system, into subensembles that display the statistics of entangled states. This motivates the introduction of the concept of "counterfactual" entanglement, which can be associated with all separable mixed states including those that are factorable. This type of entanglement gives rise to a new kind of postselection-induced Bell inequality violation. The significance of counterfactual entanglement, and its physical implications, are assessed.

It is well known that all entangled pure quantum states imply the presence of nonlocal correlations, because any such state must violate some Bell inequality [1]. For pure product states, however, no Bell inequality violation is possible, and so for these states there is no implication of nonlocal correlations. As far as *mixed* states are concerned, we can draw a distinction between "separable" states, each of which has a density matrix that can be represented as a weighted sum of projections on product states, and those mixed states for which such a representation is impossible and from which entanglement can be distilled [2]. It is impossible to distill entanglement from separable mixed states, nor can such states directly violate any Bell inequality, and it is customary to think of the correlations in these states as involving only classical statistics and being devoid of nonlocality.

Recently, however, it was pointed out [3] that any separable density matrix may contain "hidden" entanglement in that it can always be rewritten as a sum of projections on entangled states. Thus an ensemble that, considered as a whole, displays the statistics associated with a separable density matrix, may in fact have been prepared with entangled states. Hidden entanglement can be analyzed and manipulated by considering measurements on an ancilla system which, together with the subsystems to which the separable density matrix refers, constitutes an entangled pure state. The closely related notion of "entanglement of assistance" has been developed independently [4]. In this paper we introduce a new property which we call "counterfactual" entanglement, which again can be associated with separable mixtures but which is distinct from hidden entanglement. Counterfactual entanglement can be associated with separable mixed states where there is no explicit or hidden entanglement, but where measurements on an ancilla system, which, at the time these measurements are performed, is completely disentangled from the subsystems to which the separable mixed state refers, can facilitate a partitioning of the ensemble described by the separable mixed state into subensembles, each of which displays the statistics of an entangled state. It is possible to argue counterfactually that, had we carried out a Bell operator basis [5] type measurement at an earlier time on any one of these subensembles, we would with certainty have found the subensemble to be in a specific entangled state. Remarkably, this analysis can be applied with equal validity to *factorable* states, with density matrices of the form $\rho_{12} = \rho_1 \rho_2$, where the constituent subsystems do not share any entanglement with an extraneous system and need never have interacted with each other. These processes can be seen to give rise to a new kind of postselection-induced Bell inequality violation. Our results also suggest that nonlocal correlations in quantum systems may be rather more widespread than is generally thought.

Before we can introduce the idea of counterfactual entanglement, we must first make clear exactly what is meant, physically, by the term "mixed state" in quantum mechanics. The interpretation of mixed states has been the source of much confusion and debate amongst physicists (see for example [6, 7]). Mathematically, of course, we can unambiguously identify a necessary condition for a state to be mixed, by referring to its density matrix $\rho$ and using the condition $\rho^2 \neq \rho$. Physically, however, we must distinguish between the "ignorance" interpretation, according to which a mixed state simply represents a statistical mixture of individual systems each of which is in a definite pure state, and the "ancilla" interpretation, according to which a given mixed state is seen as originating in entanglement with an ancilla system, where we ignore, i.e. trace out, the ancilla. Here we take the view that the ignorance interpretation for mixed states is unsatisfactory, and that consequently the only viable interpretation for a "genuine" mixed state is based on the assumed existence of an extraneous ancilla, with which the system referred to in the mixed state is entangled.

There are several reasons for rejecting the ignorance interpretation. First of all, we can see that this interpretation is incompatible with the standard definition of a quantum state as "the most complete possible description of the state of a system". Clearly, adopting the ignorance interpretation for a particular mixed state would imply that that state was *not* the most complete possible description of the system or ensemble to which it referred. By implication, there exists in such a case *classical information* (of which we happen to be, perhaps temporarily, ignorant), which would enable us to refine our description to one that refers to a specific collection of pure states. It is, in our view, essential that when we assign a quantum state, which we take to be a fundamental description of reality, to a system, we assume that *all* exising classical information (which we take to be a higher-level and not a fundamental property) may at some point become available to us. In other words, a formulation of quantum mechanics that purports to represent it as a fundamental theory should not conflate two different interpretations of a quantum state where one of these implicitly refers to a secondary higher-level property such as classical information and the other does not.

Another reason for calling into question the validity of the ignorance interpretation is that in some cases different statistical mixtures of pure states may appear to be representable by the same density matrix but may nevertheless be experimentally distinguishable. For example, a large ensemble of spin-1/2 particles of which exactly half are prepared spin-up in the *z* direction and half are prepared spin-down in the *z* direction can be experimentally distinguished from a similar ensemble in which half the particles are prepared spin-up in the *x* direction and half spin-down in the *x* direction, even though these enembles are ostensibly describable by the same density matrix. They can be distinguished by taking a long series of measurements of spin components for each ensemble in, say, the *z* direction, and then comparing standard deviations. Similarly, genuine mixed states can in some cases be experimentally distinguished from "ignorance" mixtures [8].

It is also clear that, if we accept the validity of Bell's theorem, then the ignorance and ancilla interpretations are not compatible with each other. For example given an EPR spin singlet pair, each separate particle can be described by a mixed state, with the other particle then taking the role of ancilla. But if we then assume that this mixed state can also be given an ignorance interpretation, inconsistencies immediately arise, because this would imply that each separate particle had a definite spin-component value, thus permitting a local realistic interpretation for the EPR state, which could not be consistent with the Bell inequality violations obtainable from this state [9]. Given that the ignorance and ancilla interpretations are incompatible with each other , and that the ancilla interpretation is "democratic" in the sense that it encompasses all possible decompositions, whereas the ignorance interpretation necessarily singles out one particular decomposition, the ancilla interpretation seems preferable on grounds of generality alone.

Furthermore, it has recently been argued [6] that the density matrix can be associated with each *individual* system in an ensemble, because of the theoretical feasibility of verifying the state of an individual system by means of a "protective" measurement. This would indicate another method by which mixed states derived from entanglement with an ancilla can in principle be experimentally distinguished from ignorance mixtures with the same density matrix. This emphasizes the distinction between, on the one hand, ensembles in genuine mixed states, i.e. those derived from entanglement with an ancilla, and, on the other, ensembles corresponding to the ignorance case where each individual system is in a definite pure state.

To bring out this distinction, ensembles of the latter type have been described variously as "compounds" [10], "proper mixtures" [8], and "pseudomixed states" [3]. We conclude that genuine mixed states should always be associated with entanglement with an extraneous system. Thus any separable mixed state of two subsystems is in fact derived from an entangled three-subsystem pure state. Although the third subsystem can be thought of as fictitious for the purposes of deriving new decompositions of a given two-subsystem mixed state [3], it is also the case that, given that they are in a mixed state, there must physically exist some extraneous system with which the two subsystems are entangled.

Having established what is meant physically be a mixed state, we are now in a position to illustrate the notion of counterfactual entanglement. We consider first a system of two spin-1/2 particles which is in the separable mixed state $\rho_{12}$, where

$$\rho_{12} = \frac{1}{2}\left(|\uparrow_{1z} \uparrow_{2z}\rangle\langle\uparrow_{1z} \uparrow_{2z}| + |\downarrow_{1z} \downarrow_{2z}\rangle\langle\downarrow_{1z} \downarrow_{2z}|\right), \qquad (1)$$

which we suppose is derived from an entangled "GHZ"-type [11] pure state $|\psi_{123}\rangle$ of three spin-1/2 particles given by

$$|\psi_{123}\rangle = \frac{1}{\sqrt{2}}\left(|\uparrow_{1z} \uparrow_{2z} \uparrow_{3z}\rangle + |\downarrow_{1z} \downarrow_{2z} \downarrow_{3z}\rangle\right). \qquad (2)$$

If we carry out a spin-component measurement on particle 3, with respect to any direction except the *z*-direction [12], we can prepare particles 1 and 2 in a pseudomixed state corresponding to a specific decomposition of entangled states. This is an example of the sort of process described in [3] and [4].

However, suppose now that, before performing any measurement on particle 3, we perform a set of spin-component measurements on particles 1 and 2 with respect to the directions $\theta_1$ and $\theta_2$ respectively. There cannot then be any hidden entanglement or entanglement of assistance associated with particles 1 and 2; these particles remain in a genuine separable mixed state until the measurements of $\sigma_{1\theta_1}$ and $\sigma_{2\theta_2}$ are carried out, after which each individual pair of particles will be in a definite pure product state. Now suppose that, at a time subsequent to the $\sigma_{1\theta_1}$ and $\sigma_{2\theta_2}$ measurements, when the state of particle 3 is completely disentangled from that of particles 1 and 2, we measure $\sigma_{3\theta_3}$. We can then separate the ensemble of particles 1 and 2 into two subensembles, according to the results of each $\sigma_{3\theta_3}$ measurement; that is, we postselect each pair of particles 1 and 2 into one of two subensembles, according to whether we obtain the result +1 or -1 for the corresponding $\sigma_{3\theta_3}$ measurement. What can we then say about the statistics of the earlier $\sigma_{1\theta_1}$ and $\sigma_{2\theta_2}$ measurement results for each of these subensembles?

In fact, the distribution of results in each of these postselected subensembles will be indistinguishable from the distributions in the corresponding *pre*selected subensembles that could have been prepared by measuring $\sigma_{3\theta_3}$ *before* the measurements of $\sigma_{1\theta_1}$ and $\sigma_{2\theta_2}$ are carried out and then, subsequent to these measurements, separating the outcomes of the $\sigma_{1\theta_1}$ and $\sigma_{2\theta_2}$ measurements into two subensembles contingent on the outcomes of the earlier $\sigma_{3\theta_3}$ measurement. In the latter case, it is evident that, as we have already mentioned, each subensemble will display the statistics of an entangled state. For example, suppose that $|\uparrow_z\rangle = \alpha|\uparrow_{\theta_3}\rangle + \beta|\downarrow_{\theta_3}\rangle$. Then, in the period in between the $\sigma_{3\theta_3}$ measurement and the $\sigma_{1\theta_1}$ and $\sigma_{2\theta_2}$ measurements the whole ensemble of particles 1 and 2 will be in a pseudomixed state, and the two preselected subensembles will [3] be in the entangled pure states $\frac{1}{\sqrt{2}}\left(\alpha|\uparrow_{1z}\uparrow_{2z}\rangle + \beta^*|\downarrow_{1z}\downarrow_{2z}\rangle\right)$ and $\frac{1}{\sqrt{2}}\left(\beta|\uparrow_{1z}\uparrow_{2z}\rangle - \alpha^*|\downarrow_{1z}\downarrow_{2z}\rangle\right)$ according to whether we obtain +1 or -1 respectively for the $\sigma_{3\theta_3}$ measurement. This can be seen by rewriting $|\psi_{123}\rangle$ as

$$|\psi_{123}\rangle = \frac{1}{\sqrt{2}}\left\{|\uparrow_{3\theta_3}\rangle\left(\alpha|\uparrow_{1z}\uparrow_{2z}\rangle + \beta^*|\downarrow_{1z}\downarrow_{2z}\rangle\right) + |\downarrow_{3\theta_3}\rangle\left(\beta|\uparrow_{1z}\uparrow_{2z}\rangle - \alpha^*|\downarrow_{1z}\downarrow_{2z}\rangle\right)\right\}. \quad (3)$$

That the statistics of the $\sigma_{1\theta_1}$ and $\sigma_{2\theta_2}$ measurement results for the respective postselected subensembles should be identical to those of the preselected ones is perhaps not immediately obvious, but can easily be seen by considering the Bayesian relation:

$$\text{Prob}_{|\psi_{123}\rangle}\left(\sigma_{1\theta_1} \otimes \sigma_{2\theta_2} = j|\sigma_{3\theta_3} = i\right) = \frac{\text{Prob}_{|\psi_{123}\rangle}\left(\sigma_{1\theta_1} \otimes \sigma_{2\theta_2} = j\right)\text{Prob}_{|\psi_{123}\rangle}\left(\sigma_{3\theta_3} = i|\sigma_{1\theta_1} \otimes \sigma_{2\theta_2} = j\right)}{\text{Prob}_{|\psi_{123}\rangle}\left(\sigma_{3\theta_3} = i\right)}$$
(4)

In eq. (4) the left hand side represents the preselected case, i.e. the probability for obtaining the result $\sigma_{1\theta_1} \otimes \sigma_{2\theta_2} = j$ given that the outcome $\sigma_{3\theta_3} = i$ has already been obtained, whilst the right hand side represents the postselected case where the order of the $\sigma_{1\theta_1} \otimes \sigma_{2\theta_2}$ and $\sigma_{3\theta_3}$ measurements is reversed.

The equivalence of the statistics of the preselected and postselected subensembles can also be seen by considering the case where the $\sigma_{3\theta_3}$ measurement is spacelike separated from the measurements of $\sigma_{1\theta_1}$ and $\sigma_{2\theta_2}$. In this case different Lorentz observers could interpret the contingent subensembles of particles 1 and 2 as preselected or postselected by the $\sigma_{3\theta_3}$ measurement, depending on their state of motion. There would thus be a serious violation of Lorentz invariance, at the observable level, if the preselected and postselected subensembles did not yield identical statistics.

The foregoing analysis shows that the postselected subensembles of particles 1 and 2 must display the statistics of entangled states, even though they are at all times actually in separable states. We can therefore say that each postselected subensemble has *counterfactual* entanglement associated with it. Counterfactual entanglement, like actual entanglement, implies that a Bell inequality violation is possible. For example, if we carry out a Bell inequality testing experiment by performing a series of measurements of $\sigma_{1\theta_1}, \sigma_{2\theta_2}, \sigma_{1\phi_1}$ and $\sigma_{2\phi_2}$ in the usual way, then we will not be able to obtain any Bell inequality violation for the whole ensemble described by the initial state $|\psi_{123}\rangle$. But if at a later stage we measure $\sigma_{3\theta_3}$ and segregate the earlier results of the $\sigma_{1\theta_1}, \sigma_{2\theta_2}, \sigma_{1\phi_1}$ and $\sigma_{2\phi_2}$ measurements into two subensembles according to whether $\sigma_{3\theta_3} = \pm 1$, then each subensemble will be able to yield a Bell inequality violation for suitable choices of $\theta_{1,2}$ and $\phi_{1,2}$.

This analysis can straightforwardly be extended to any separable mixed state: hence all separable mixed states must incorporate counterfactual entanglement. Thus it could be argued that all separable mixed states must be nonlocally correlated. Even if we do not have access to the extraneous "traced out" system associated with a given separable mixed state, the fact that this extraneous system must exist means that it will always be possible *in principle* to partition the outcomes of individual measurements on any separable mixed state into sets corresponding to counterfactually entangled subensembles, by measuring a suitable observable of the extraneous system at some later stage. Although this partitioning may be very difficult to implement in practice, the fact that it is in principle always possible is sufficient to imply the presence of nonlocal correlations within all systems described by separable mixed states. It should be emphasized that such a partitioning will in principle always be feasible. There is no fundamental reason why the necessary measurements on the ancilla could not be carried out, even though they may be extremely difficult to implement.

Remarkably, the above analysis applies with equal validity to the special case of *factorable* mixed states, i.e. those states with density matrices of the form $\rho_{12} = \rho_1 \rho_2$, where the constituent subsystems do not jointly share any entanglement with an extraneous system and need never have interacted with each other. Suppose, for example, two spin-1/2 particles are described by the factorable mixed state $\rho_{12}$ given by

$$\rho_{12} = \rho_1 \rho_2 = \frac{1}{2}\left(|\uparrow_{1z}\rangle\langle\uparrow_{1z}| + |\downarrow_{1z}\rangle\langle\downarrow_{1z}|\right)\frac{1}{2}\left(|\uparrow_{2z}\rangle\langle\uparrow_{2z}| + |\downarrow_{2z}\rangle\langle\downarrow_{2z}|\right). \qquad (5)$$

We assume that $\rho_1$ is derived from an entangled pure state involving a third spin-1/2 particle so that $\rho_1 = Tr_3\left(|\psi_{13}\rangle\langle\psi_{13}|\right)$, where $|\psi_{13}\rangle = \frac{1}{\sqrt{2}}\left(|\uparrow_{1z} \uparrow_{3z}\rangle + |\downarrow_{1z} \downarrow_{3z}\rangle\right)$, and that similarly $\rho_2$ is derived from the pure state $|\psi_{24}\rangle$ so that $\rho_2 = Tr_4\left(|\psi_{24}\rangle\langle\psi_{24}|\right)$

where $|\psi_{24}\rangle = \frac{1}{\sqrt{2}}\left(|\uparrow_{2z} \uparrow_{4z}\rangle + |\downarrow_{2z} \downarrow_{4z}\rangle\right)$. Thus particles 1 and 2 do not jointly share any entanglement with either of the ancilla particles 3 and 4, and the four-particle pure state from which $\rho_{12}$ is derived can be written as

$$|\psi_{1234}\rangle = \frac{1}{2}\left(|\uparrow_{1z} \uparrow_{2z} \uparrow_{3z} \uparrow_{4z}\rangle + |\uparrow_{1z} \downarrow_{2z} \uparrow_{3z} \downarrow_{4z}\rangle + |\downarrow_{1z} \uparrow_{2z} \downarrow_{3z} \uparrow_{4z}\rangle + |\downarrow_{1z} \downarrow_{2z} \downarrow_{3z} \downarrow_{4z}\rangle\right)$$
(6)

Now, if we carried out a Bell operator basis measurement on particles 3 and 4, it would be possible to prepare particles 1 and 2 in one of the entangled Bell operator eigenstates; this would be an elementary example of "entanglement swapping" [13]. However, suppose instead that we perform local measurements of $\sigma_{1\theta_1}$ and $\sigma_{2\theta_2}$ before any measurement on particles 3 and 4 is carried out. Then, when particles 3 and 4 are fully disentangled from particles 1 and 2, we can carry out a Bell operator basis measurement on particles 3 and 4, the outcome of which will enable us to postselect Bell inequality violating subensembles of particles 1 and 2, as in the previous example. What is striking about the current example is the implication of nonlocal correlations in the postselected subensembles of particles 1 and 2, even though these particles remain in a factorable state throughout the process. In the previous (GHZ) example one might attempt to explain the apparent nonlocal correlations between particles 1 and 2 as arising from their shared entanglement with a third system; but in the factorable case this explanation will not get off the ground.

In assessing the significance of counterfactual entanglement, it is worth bearing in mind that standard quantum mechanics does not allow one to make counterfactual inferences about the earlier states of quantum systems, based on the outcomes of subsequent measurements. Thus, according to standard quantum mechanics, we cannot for example argue that a system prepared at time $t_0$ in the state $|\psi_{123}\rangle$ given by eq. (2), subjected to measurements of $\sigma_{1\theta_1}$ and $\sigma_{2\theta_2}$ at time $t_1$, and then postselected by the outcome $\sigma_{3x} = 1$ at time $t_2$ (where $t_0 < t_1 < t_2$) would, if it had been subjected to a Bell operator basis measurement at time $t_{1/2}$ (where $t_0 < t_{1/2} < t_1$), have necessarily yielded the eigenstate $\frac{1}{\sqrt{2}}\left(|\uparrow_{1z} \uparrow_{2z}\rangle + |\downarrow_{1z} \downarrow_{2z}\rangle\right)$, even though the postselected subensemble yields identical statistics to those that would have been obtained for this eigenstate, for any choice of $\theta_1$, $\theta_2$. In this sense standard quantum mechanics is predictive but not retrodictive.

However, there has recently been a good deal of interest in the possibility of making retrodictive inferences of this kind, particularly within the context of the consistent histories [14] interpretation of quantum mechanics and related time-symmetric interpretations (see for example [15]). The consistent histories interpretation requires that consistency conditions be satisfied in any given example before classical probabilities can be assigned to quantum events, and it has been suggested [16] that similar conditions must be satisfied by related interpretations in order to preclude the possibility of a direct contradiction with standard quantum mechanical predictions [17]. In general these conditions can be written as

$$\mathrm{Re}\,\mathrm{Tr}\left(\hat{E}_n \hat{E}_{n-1} \cdots \hat{E}_k^\alpha \cdots \hat{E}_1 \hat{D} \hat{E}_1 \cdots \hat{E}_k^\beta \cdots \hat{E}_{n-1} \hat{E}_n \hat{F}\right) = 0 \qquad (7)$$

for every pair $\alpha < \beta$, where the projection operators $\hat{E}_i$ refer to series of events occurring in between the initial and final events $\hat{D}$ and $\hat{F}$ respectively. Provided these conditions are satisfied, the expression

$$\frac{\mathrm{Tr}\left(\hat{E}_n \hat{E}_{n-1} \cdots \hat{E}_2 \hat{E}_1 \hat{D} \hat{E}_1 \hat{E}_2 \cdots \hat{E}_{n-1} \hat{E}_n \hat{F}\right)}{\mathrm{Tr}\left(\hat{D}\hat{F}\right)} \tag{8}$$

can be understood as the classical probability $\mathrm{Prob}(E_1 \wedge E_2 \wedge \cdots \wedge E_n | D \wedge F)$. The related idea of a "consistent framework" has also been proposed [18].

Although it has given rise to a number of conceptual difficulties [8], the consistent histories approach can be applied to assess the validity of counterfactual retrodictive inferences [19] of the type we have considered. In fact we find that, for the counterfactual entanglement examples we have looked at, the consistency conditions given by eq. (7) are satisfied and the probability associated with the projection on the relevant entangled state prior to the $\sigma_{1\theta_1}$, $\sigma_{2\theta_2}$ measurements, as given by (8), is unity in each case. For example, in the case of the state given by eq.s (1) and (2), if we postselect by $\sigma_{3x} = 1$ after earlier measurements of $\sigma_{1\theta_1}$ and $\sigma_{2\theta_2}$ we can write $\hat{D} = |\psi_{123}\rangle\langle\psi_{123}|$, $\hat{F} = |\sigma_{3x} = 1, \sigma_{1\theta_1} = i, \sigma_{2\theta_2} = j\rangle\langle\sigma_{3x} = 1, \sigma_{1\theta_1} = i, \sigma_{2\theta_2} = j|$, and then consider a set of possible Bell operator basis projections at time $t$ in between the initial time associated with $\hat{D}$ and the time of the $\sigma_{1\theta_1}$, $\sigma_{2\theta_2}$ measurements. We find that the conditions given by eq. (7) are satisfied and the expression (8) yields probability 1 for the projection on the counterfactual entangled state $\frac{1}{\sqrt{2}}\left(|\uparrow_{1z} \uparrow_{2z}\rangle + |\downarrow_{1z} \downarrow_{2z}\rangle\right)$ at time $t$. A similar result is obtained in the factorable case (eq.s (5) and (6)), where we postselect by a Bell operator measurement yielding the state $\frac{1}{\sqrt{2}}\left(|\uparrow_{3z} \uparrow_{4z}\rangle + |\downarrow_{3z} \downarrow_{4z}\rangle\right)$ for particles 3 and 4.

The fact that the postselected subensembles we have looked at can in some cases give maximal violation of Bell inequalities strengthens the case for attributing counterfactual properties, in a retrodictive sense, to quantum systems. If we do not accept that the relevant postselected subensemble would have yielded a specific maximally entangled eigenstate had it been subjected to a Bell operator basis type measurement at an earlier time, then how can we explain the maximal Bell inequality violation displayed by that subensemble?

It is interesting to compare the role of postselection in the processes we have described with previous examples of deriving Bell inequality violations via postselection [20-23]. It is well known that, if we disallow certain ranges of measurement outcomes, which are then discarded, then even classical physics can produce Bell inequality violations. An example of this type (which does not of course imply the presence of any kind of nonlocal correlation) is described in [20]. This type of process can be described as *measurement-dependent* postselection, since the subensembles selected may depend on the measurement carried out; that is, the discarded outcomes may be biased towards a particular measurement or set of measurements. Similarly, Bell inequality violations via measurement-dependent postselection can be demonstrated for separable mixed states in quantum mechanics [21, 22].

A different kind of postselection-induced Bell inequality violation can be demonstrated for the Werner states [24], which do not violate any Bell inequality for single ideal measurements but can nevertheless violate Bell inequalities if a subensemble is postselected according to the results of one measurement and this subensemble is then subjected to another measurement [23]. This type of postselection could be described as a *nonlocal* selection process, in that the postselection involves comparing and combining two sets of results that occur at spatially separated locations. Examples involving this type of postselection are interesting even though the states they refer to, such as the Werner states, are not separable; the Bell inequality violations generated are nevertheless postselection-induced.

The role of postselection in the examples we have analyzed in this paper is different again. In the counterfactual entanglement examples no results are disallowed, and so there is no measurement-dependent selection. Also, only single ideal measurements are performed on the relevant subensembles, and the postselection is local, that is, it does not involve any nonlocal selection process. The postselection is carried out, as we have seen, via local measurements on an extraneous system.

An analysis of possible applications of the counterfactual entanglement concept would be beyond the scope of this paper. It is clear, however, that any quantum information processing application that uses entanglement as a resource can in principle be carried out with separable states. But if such a method is used, measurements on the disentangled extraneous system, in an appropriate basis, will have to be performed in order to partition the counterfactually entangled subensembles correctly. In this way ultimate control of each application could be channeled to a remote location and the final decision to implement the application could be delayed until any chosen time subsequent to the main processing itself.

In view of the arguments we have presented, it is worth considering whether it is appropriate to label weighted sums of projections on product states as "separable" mixed states, as is the current convention. The usual rationale for such labeling is that mixtures of this kind can, it is claimed, be prepared by separated experimenters receiving instructions from a central source [25] or exchanging information with each other [26]. However, it is evident that, as far as genuine mixed states are concerned, only those separable mixed states that are factorable can in fact be prepared by such a method. Factorable states apart, only *pseudo*mixed states, which are really collections of pure states and as such are in some cases experimentally distinguishable from the corresponding genuine mixed states [6, 8], can be prepared in this way. Since, as we have seen, all separable mixed states, including the factorable ones, incorporate counterfactual entanglement implying the presence of nonlocal correlations, the conventional labeling is perhaps a source of potential confusion.

In closing, we can see that it is only the pure product states that are devoid of actual or counterfactual entanglement, and hence, by implication, of nonlocal correlations. However, it could be argued that in practice pure states are almost always unrealistic idealizations, because of the impossibility of completely screening off environmental interactions. Assuming that Bell inequality violations of the type we have described constitute definite evidence of nonlocality, one would then be led to the conclusion that nonlocal correlations are ubiquitous in the physical world. Alternatively, one could question the validity of Bell inequality violations as genuine evidence of nonlocality [27]; our results could then be seen as emphasizing the dangers of accepting too readily the view that Bell inequality violations can be interpreted in that way.

The author is grateful for support from the Leverhulme Trust.